\documentclass[prd,nofootinbib,preprintnumbers,twocolumn,
showpacs,floatfix]{revtex4-2}
\usepackage[english]{babel}

\usepackage[letterpaper,top=2cm,bottom=2cm,left=1.5cm,right=1.5cm,marginparwidth=1.75cm]{geometry}
\usepackage{caption}
\usepackage{subfig}
\usepackage{amsmath}
\usepackage{graphicx}
\usepackage[colorlinks=true, allcolors=blue]{hyperref}

\usepackage{float}
\usepackage{booktabs}
\usepackage{color}

\newcommand{\be}{\begin{equation}}
\newcommand{\ee}{\end{equation}}
\newcommand{\bea}{\begin{eqnarray}}
\newcommand{\eea}{\end{eqnarray}}

\newcommand{\WJ}[1]{\textcolor{cyan}{#1}}

\begin{document}
\title{Detecting gravitational waves with spin systems}

\author{Jiamin Liang$^1$}
\author{Mingqiu Li$^1$}
\email[]{limingqiu17@mails.ucas.ac.cn}
\author{Yu Gao$^{2,3}$}
\email[]{gaoyu@ihep.ac.cn }

\author{Wei Ji$^5$}
\email[]{wei.ji@pku.edu.cn}

\author{Sichun Sun $^1$}
\email[]{sichunssun@gmail.com}
\author{Qi-Shu Yan$^{2,4}$}
\email[]{yanqishu@ucas.ac.cn}

\affiliation{$^1$School of Physics, Beijing Institute of Technology, Beijing, 100081, China}
\affiliation{$^2$Institute of High Energy Physics, Chinese Academy of Sciences, Beijing 100049, China}
\affiliation{$^3$State Key Laboratory of Particle Astrophysics, Institute of High Energy Physics, Chinese Academy of Sciences, Beijing 100049, China}
\affiliation{$^4$University of Chinese Academy of Sciences, Beijing 100049, China}
\affiliation{$^5$School of Physics and State Key Laboratory of Nuclear Physics and Technology, Peking University, Beijing 100871, China}

\begin{abstract}
The observation of gravitational waves has opened a new window into the Universe through gravitational-wave astronomy. However, high-frequency gravitational waves remain undetected. In this work, we propose that spin systems can be employed to detect gravitational waves in this unexplored frequency regime. We derive the spin’s response to gravitational waves and identify three distinct effects: the well-known Gertsenshtein effect, a metric-induced interaction, and the gravitational spin Hall effect. We focus on nuclear spins and utilize nuclear magnetic resonance to enhance the gravitational response, leveraging the advantages of long coherence time, high polarization, and a small gyromagnetic ratio. The proposed experimental scheme is capable of probing gravitational waves in the kilohertz to gigahertz range, with projected sensitivities reaching $\sqrt{S_h}\approx10^{-20}~\mathrm{Hz}^{-1/2}$. 

\end{abstract}
\maketitle
{\bf Introduction.}~
The discovery of gravitational waves (GW)~\cite{LIGOScientific:2016aoc} marks the beginning of an era in which the Universe can be observed with an essential new tool, complementing electromagnetic waves. Analogous to electromagnetic waves, which reveal diverse phenomena across a wide range of wavelengths, gravitational waves encompass significant physics across a wide range of frequencies, highlighting the importance of studying each spectral band. There are current and future projects, such as ground-based and space-borne interferometers~\cite{Guo:2018npi,Luo:2015ght,Audley:2017drz, Cornish:2018dyw,LIGOScientific:2016aoc} and pulsar timing arrays~\cite{Xu:2023wog,NANOGrav:2023hvm,NANOGrav:2023gor,EPTA:2023fyk} at low frequencies. For frequencies higher than kilohertz, several proposals~\cite{Gao:2023gph,Berlin:2021txa,Ito:2022rxn,Gao:2023ggo,Kanno:2023whr} aim at astrophysical sources.

Two main approaches have been developed to study the interaction between spin and gravity. One was generalizing the Dirac equation to curved space-time, using the Foldy-Wouthuysen transformation to obtain a non-relativistic  theory~\cite{Obukhov:2000ih}.
The second approach was the pure classical approach, mainly based on the Mathisson-Papapetrou-Dixon equations~\cite{Costa:2017kdr} or the classical approximation of the Dirac equation in curved spacetime~\cite{Obukhov:2009qs}. In ~\cite{Ruggiero:2020oxo,Ruggiero_2020}, the effects of a gravitational plane wave on particle spin are described in terms of a gravito-electromagnetic analogy, which implies that magnetometers can detect GWs. For a charged fermion in curved space-time under a magnetic field background, the non-relativistic Hamiltonian has been derived~\cite{Asprea:2020fuy,Ito:2020wxi}. Magnons or other electromagnetic spin-related effects can be excited by GWs in a ferromagnetic or antiferromagnetic sample as in~\cite{Ito:2020wxi,Ito:2022rxn, Lei:2025vek}. Magnon GW detectors were proposed in~\cite{Ito:2020wxi,Ito:2022rxn}, and an upper limit on GHz GW was obtained by reinterpreting the existing data from axion dark matter experiments. However, we should note that in~\cite{Ito:2020wxi,Ito:2022rxn} the Gertsenshtein effect was overlooked, which we will show can dominate for a spin system. 

We propose using nuclear magnetic resonance (NMR) as a novel method for detecting gravitational waves. By tuning the applied magnetic field, the resonance frequency of nuclear spins can be precisely controlled, enabling gravitational wave searches in frequency ranges that are inaccessible to conventional techniques such as cavity-based detectors. Moreover, nuclear spins offer key advantages, including long coherence times and high polarization, which can significantly amplify the signal strength and thus enhance detection sensitivity. A similar scheme has been successfully employed in searches for dark matter and new physics beyond the Standard Model, such as the CASPEr experiment~\cite{budker2014proposal,garcon2017cosmic,wei2025dark} and exotic spin-dependent force experiments~\cite{gkika2024optimization,arvanitaki2014resonantly,xu2025constraints,heng2025search}.


In this work, we begin with the Dirac equation in a curved spacetime background, then identify three distinct effects: the well-known Gertsenshtein effect, a previously overlooked metric-induced effect, and the gravitational spin Hall effect. The key conclusion is that, besides the Gertsenshtein effect, fermion spins are also sensitive to metric perturbation, and both of these effects have been sometimes overlooked in prior literature on spin-based setups.
Then we explore the detectability of gravitational waves using spin-based sensors. We demonstrate that NMR techniques can amplify the gravitational-wave-induced signal by up to six orders of magnitude 
, enabling high-sensitivity detection in the MHz frequency range. This approach is particularly promising for probing frequency bands that are challenging for existing detection methods.
Finally, we summarize our findings and discuss future directions.

{\textbf{Fermions in curved spacetime.}\label{sec2}
To study the effects of gravity on a fermion, we consider the Dirac equation in
curved spacetime with a metric $g_{\mu\nu}$, which is
\begin{align}
 i\gamma^{\hat{\alpha}}e_{\hat{\alpha}}^{\mu}\left(\partial_{\mu}+\Gamma_{\mu}+i e A_{\mu}\right)\psi=m\psi~,   
\end{align}
where   $e_{\hat{\alpha}}^{\mu}$ is the tetrad,
\begin{align}
    e_{\mu}^{\hat{\alpha}}e_{\nu}^{\hat{\beta}}\eta_{\hat{\alpha}\hat{\beta}}=g_{\mu\nu}\,, 
\end{align}
with $\eta_{\hat{\alpha}\hat{\beta}}=\text{diag}(1,-1,-1,-1)$.
The spin connection is defined by
\begin{align}
    \Gamma_{\mu}=-\frac{i}{2}e_{\nu}^{\hat{\alpha}}\sigma_{\hat{\alpha}\hat{\beta}}\left(\partial_{\mu}e^{\nu\hat{\beta}}+\Gamma_{\lambda\mu}^{\nu}e^{\lambda\hat{\beta}}\right),
\end{align}
where $\sigma_{{\hat{\alpha}}{\hat{\beta}}}={\textstyle\frac{i}{4}}[\gamma_{{\hat{\alpha}}},\gamma_{{\hat{\beta}}}]$ is a generator of the Lorentz group and $\Gamma^\nu_{\mu\lambda}$ is the Christoffel symbol.
To isolate the effect of GWs, we linearize the metric as $g_{\mu\nu}=\eta_{\mu\nu}+h_{\mu\nu}$, where $\eta_{\mu\nu}$ is the Minkowski metric. 
In the transverse-traceless (TT) gauge, without loss of generality, $h_{ij}$ can take the form
\begin{equation}\label{eqGWS}
\begin{aligned}
 h_{ij}^{TT}=&h_+\cos(\omega t-{\bf k}\cdot {\bf x})e_{ij}^{(+)}\\
 &+ h_\times\cos(\omega t-{\bf k}\cdot {\bf x}+\alpha)e_{ij}^{(\times)}\,,  
\end{aligned}
\end{equation}
 where $\hat{k}=(\sin \theta, 0, \cos\theta)$ is the direction of GWs and $e_{i j}^{(+)}$ and $e_{i j}^{(\times)}$ are the gravitational wave basis tensors.

In Fermi normal coordinates, 
the tensor perturbations $h_{ij}$ are
\begin{equation}\label{fnhij}
 \begin{aligned}
 h_{00}&=-{ R}_{0i0j}x^{i}x^{j},\\ h_{0i}&=-\frac{2}{3}{ R}_{0j i k}x^{j}x^{k},\\ h_{i j}&=-\frac{1}{3}{ R}_{i k j l}x^{k}x^{l}\,, 
 \end{aligned}  
\end{equation}
where the Riemann tensor is evaluated at $x^i=0$, so it depends only on time. To first order in $h_{ij}$, the Riemann tensor is the same in any frame, thus we can evaluate it in TT gauge.

In Fermi normal coordinates and to first order in $h_{\mu\nu}$, the Dirac equation reduces~\cite{Ito:2020wxi} to $i\partial_{t}\psi\;=\;\hat{H}\psi$,
in which the spin-related Hamiltonian can be written with several equivalent fields,
\be 
H=-\frac{e}{2m}\sigma^i(B^i+B^i_{\rm metric}+B^i_{\rm grav.mag.}+B^i_{ \rm grav.orb.}),
\ee
where the first $B^i$ term is the standard spin-magnetic field coupling in the Pauli equation, including the Gertsenshtein effect. The other terms take the form
\be\label{eqH}
\begin{aligned}
 B^i_{\rm metric}&=\frac{m}{2e}\epsilon_{l j i}\dot{h}_{k l,j}x^{k},\\
  B^i_{\rm grav.mag.}&=B^{j}\left[-\delta_{i j}\frac{1}{3}\ddot{h}_{k l}x^{k}x^{l}-\right.\\&\left.\frac{1}{12}\left(h_{i l,k j}+h_{k j,i l}-h_{k l,i j}-h_{i j,k l}\right)x^{k}x^{l}\right],\\
    B^i_{\rm grav.orb.}&=\frac{1}{4e}\epsilon_{k j i}\left(h_{k m,j l}-h_{k l,j m}-\delta_{j m}\ddot{h}_{k l}\right)\\&\cdot x^{l}\left(i\partial_m+eA_m\right).
\end{aligned}
\ee
The first $B^i_{\rm metric}$ term is derived from the gravitational effect on the particle spin~\cite{Ruggiero_2020,Ruggiero:2020oxo}.
The following term represents interaction between gravity and a spin inside an external magnetic field, which causes spin resonance and magnon excitations~\cite{Ito:2020wxi}.
The last term is a gravity-mediated spin-orbit coupling, yielding the gravitational spin Hall effect~\cite{Wang:2023bmd}.  Note that for composite particles like atoms, $e/m$ in Eqs.(\ref{eqH}) should be replaced with the gyromagnetic ratio $\gamma$.


In addition to the external magnetic field applied to a spin, the induced magnetic field due to the Gertsenshtein effect should not be neglected. We derived in the Appendix the detailed expressions for the induced B fields above and the well-studied Gertsenshtein effect.
The typical magnitude of each term can be estimated as
\bea
B^i_{\rm grav.mag.}&\sim& h\frac{L^2}{\lambda^2}B_0, \\
B^i_{\rm metric} &\sim& h\frac{L}{\lambda^2}\frac{m}{e}, \\ B^i_{\rm grav.orb.}&\sim &h\frac{L}{\lambda^2}\frac{p}{e},\\
B^i_{\rm Gert.}&\sim& h\frac{L}{\lambda}B_0, 
\eea
where $L$ is the maximal size of the magnetic field covered region, and $\lambda$ is the GW wavelength. Please note that $B^i_{\rm grav.mag.}$ is proportional to $L^2/\lambda^2$ while $B^i_{\rm metric}$ and $B^i_{\rm grav.orb.}$ are proportional to $L/\lambda^2$, and $B_{\rm Gert.}$ is proportional to $L/\lambda$. In Fermi normal coordinates, $\lambda\gg L$ should be fulfilled, which means $B_{\rm Gert.}\gg B_{\rm grav.mag.}^i$. For non-relativistic particles, $p\ll m$, thus $B_{\rm grav.orb.}^i\ll B_{\rm metric}^i$. The $B^i_{\rm metric}$ is inversely proportional to the gyromagnetic ratio $e/m$ of the particle. The dominant effect of GWs on a spin is the Gertsenshtein effect or the $B_{\rm metric}^i$ term. For electrons, $B^i_{\rm metric}\sim 2.4\times 10^{-9}\,\text{T}\cdot h (\frac{10\text{m}}{\lambda})^2\frac{L}{0.1\text{m}}$. While the $B^i_{\rm Gert.}\sim 10^{-8}\,\text{T}\cdot h\frac{B_0}{\rm 1\mu T}$ for $L/\lambda=0.01$, and it can be much larger with larger leading field $B_0$.




The signal strength is proportional to GW $h$. For a benchmark, the GW frequency for the innermost stable circular orbit for a pair of primordial Black Hole binary with equal mass of $M_{\rm PBH}$ is $f_{\rm ISCO}\sim 2.2\text{kHz}\cdot M_\odot/M_{\rm PBH}$. The GW strain produced by the binary is $h\sim 2(GM)^{5/3}(\pi f)^{2/3}/D$, where $M$ is the binary chirp mass, $D$ is the distance to the observer. Taking $M_{\rm PBH}=10^{-3}M_\odot, D=1.6{\rm kpc} $, the frequency $f_{\rm ISCO}\sim 2.2{\rm MHz}$ and corresponding GW strain $h\sim 10^{-20}$. This gives an idea of the sensitivity requirement for potential GW detection. 
 In addition to binary systems, the superradiance system is another GW source to be detected, possibly by the spin systems.  Bosonic fields centered on a Kerr Black Hole can emit GWs by the superradiance mechanism. The GWs by superradiance mechanism is almost monochromatic with frequency $f\sim \mu/\pi\sim 4.76{\rm Hz}\frac{\mu}{10^{-14}{\rm eV}}$. The GWs train amplitude can be evaluated by \cite{Guo:2024dqd}
\begin{equation}
	\begin{aligned}
	h\approx 1.1\times 10^{-19}\frac{M_{\rm BH}}{M_\odot}\frac{\rm Mpc}{D}\frac{M_{\mathrm{v,max}}}{M_{\rm BH}}(GM_{\rm BH}\mu)^4\,,\\
	\end{aligned}   
\end{equation}
and the duration timescale
\begin{equation}\label{srtime}
	\begin{aligned}
		t_{d}\approx &2.1\,{\rm days}\\
		&\times\left(\frac{M_{\rm BH}}{10M_\odot}\right)\left(\frac{0.1}{GM_{\rm BH}\mu}\right)^{10}\left(\frac{0.1}{M_{\mathrm{v,max}}/M_{\rm BH}}\right),   
	\end{aligned}   
\end{equation}
where $M_{\mathrm{v,max}}$ is the maximum mass of bosonic cloud and $\mu$ is mass of bosonic particles. The superradiance system is an ideal GW source for detection, as it emits long-lasting, monochromatic GWs.


 Figure~\ref{fig:plota2} shows Gertsenshtein effect and $B^i_{\rm metric}$ under different background magnetic field $B_0$. The solid red and solid blue lines indicate the conditions under which the amplitude of the magnetic field induced by the Gertsenshtein effect equals $B_{\rm metric}^i$ for nuclei ($^3\text{He}$, for example) and electrons, respectively. For $^3\text{He}$ in the region below the red line, the metric-induced field $B_{\rm metric}^i$ exceeds the amplitude of the Gertsenshtein-induced field, meaning that $B_{\rm metric}^i$ dominates the spin dynamics. 
 For electrons, due to the large gyromagnetic ratio, the critical background magnetic field is much smaller than that of the $^3\text{He}$.

\begin{figure}[t]
	\includegraphics[width=0.95\linewidth]{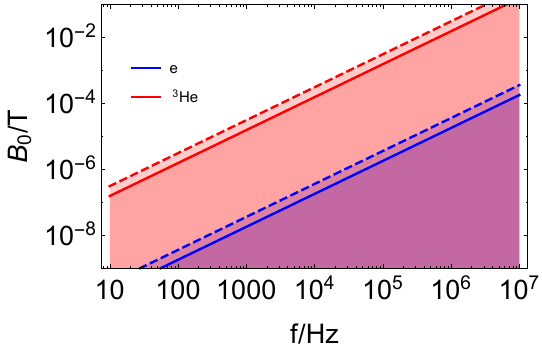}
	\caption{The solid lines show the conditions when the amplitude of the induced magnetic field by the Gertsenshtein effect is equal to that from $B^i_{\rm metric}$, the gravitational effect on spin. The signal from the equivalent field $B^i_{\rm metric}$ can be more significant with a lower external $B_0$ and a smaller gyromagnetic factor. 
	The dashed lines are the dependence relations between the external magnetic field and the corresponding Larmor frequency of electron and ($^3{\rm He}$) nucleus. 
 }
	\label{fig:plota2}
\end{figure}

The frequency of a spin system, particularly in magnetic resonance-based magnetometry, follows the Larmor precession relation: $f_L = \gamma B$, where $\gamma$ is the gyromagnetic ratio of the fermion. For example, $\gamma_e \approx 2.8 \times 10^{10} \,\mathrm{Hz/T}$ for electron and $\gamma_{\mathrm{He}} \approx 32.4 \,\mathrm{MHz/T}$ for $^3$He, $\gamma=74.5$MHz/T for $^{129}$Xe nuclei respectively\,\cite{mohr2025codata}. In Fig.\ref{fig:plota2}, the applied magnetic field and the corresponding Larmor frequencies are shown in dashed lines. Notice that both the critical frequency of GWs and the Larmor precession increase when the background field $B_0$ increases.
 Conventional magnetic resonance magnetometers operate near their resonance frequencies (in a near-resonance mode). Though some designs operate at frequencies away from resonance, magnetometer sensitivity typically drops rapidly, especially at frequencies above resonance. The dashed lines lie slightly above the solid lines, indicating that the Gertsenshtein effect slightly dominates over the metric-induced effect when using a magnetic resonance magnetometer. Nevertheless, the contribution from $B_{\rm metric}^i$ remains significant and cannot be neglected, as it plays an essential role in the overall magnetic response.

 \begin{figure}[hbt]
		\subfloat[$B_{\rm metric}^i$  ($h_+$).]{\includegraphics[width=0.45\linewidth]{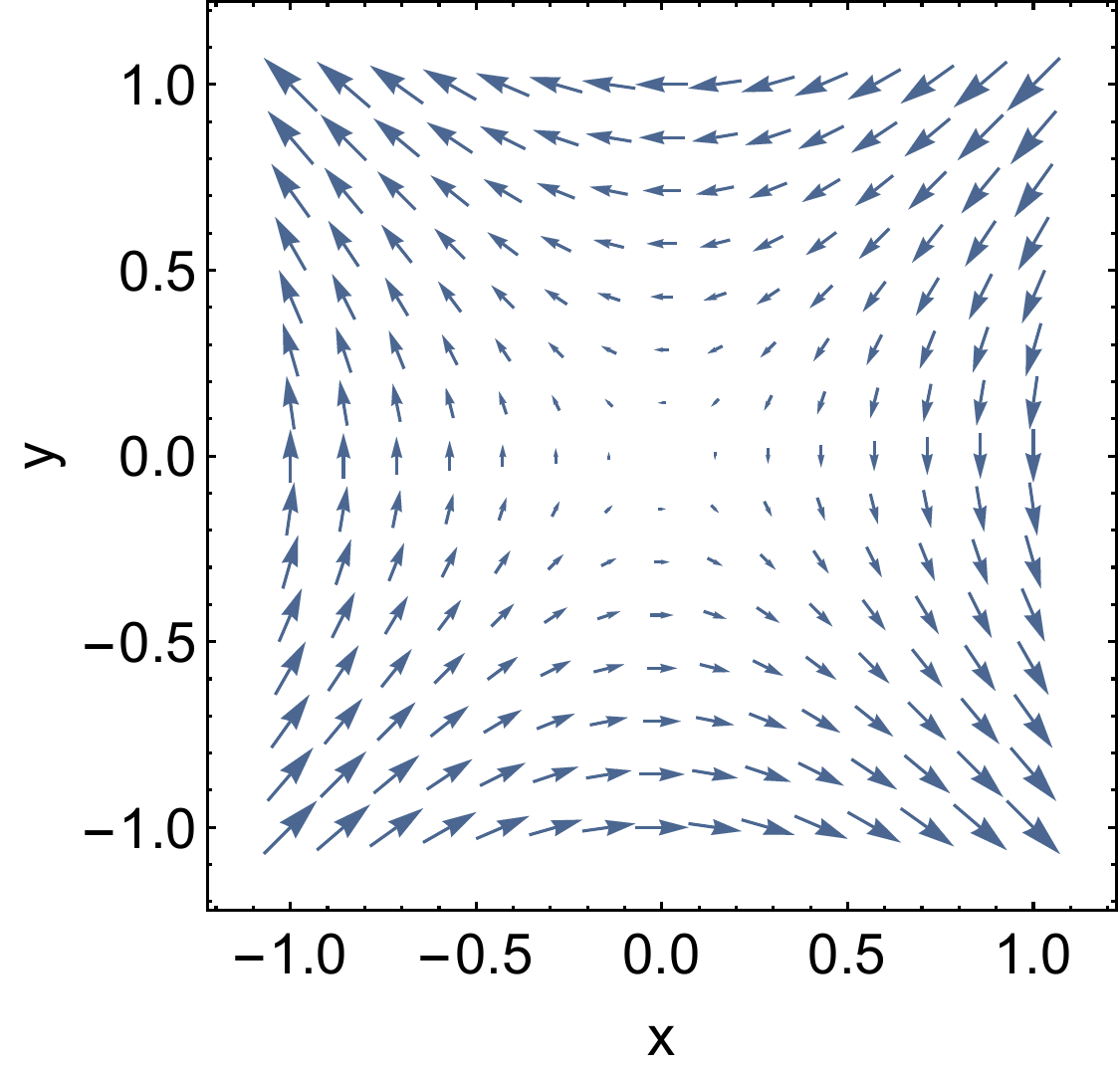} }
		\subfloat[$B_{\rm metric}^i$  ($h_\times$).]{\includegraphics[width=0.45\linewidth]{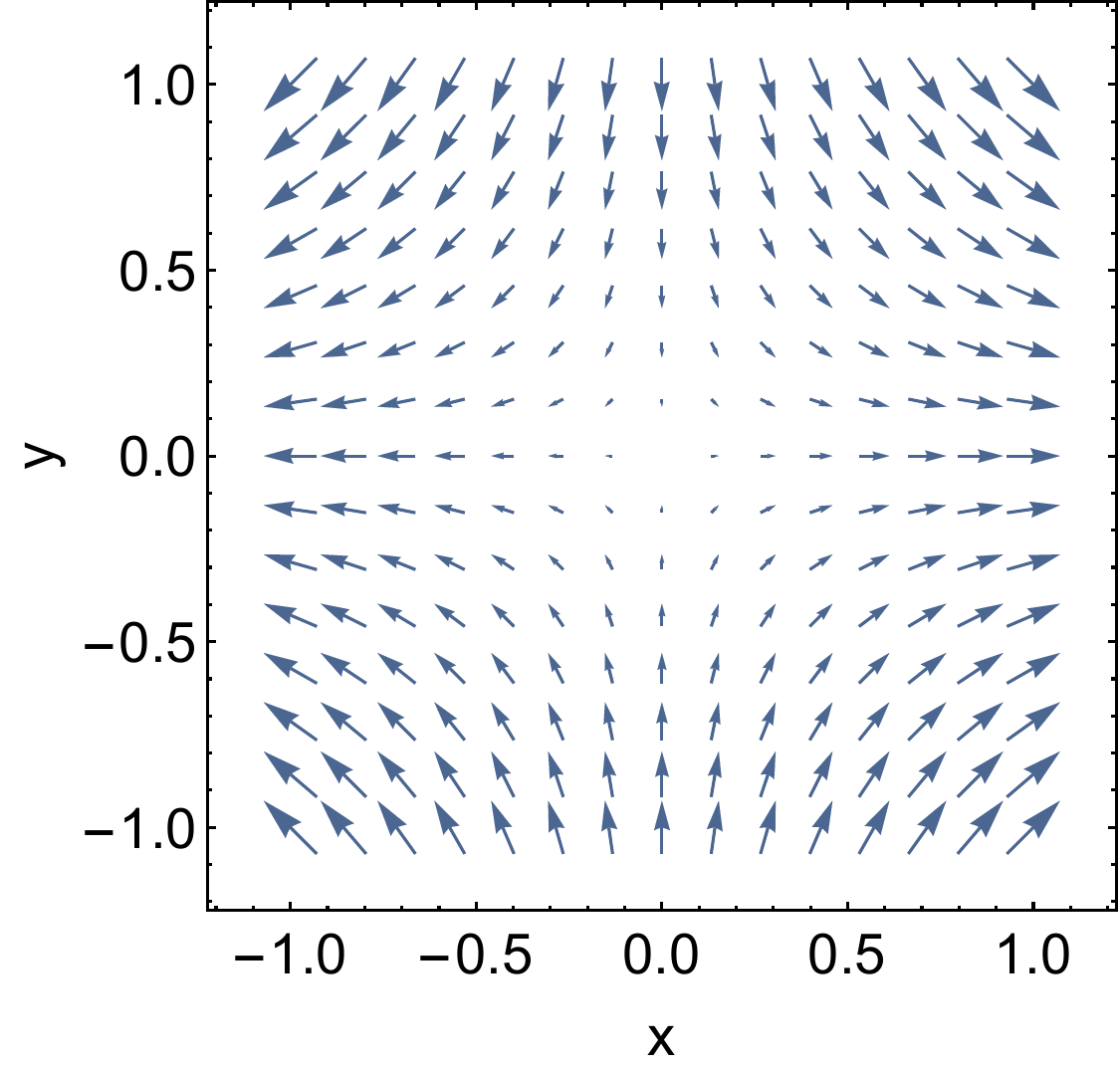}}\quad
		\subfloat[Gert. effect ($h_+$).]{\includegraphics[width=0.52\linewidth]{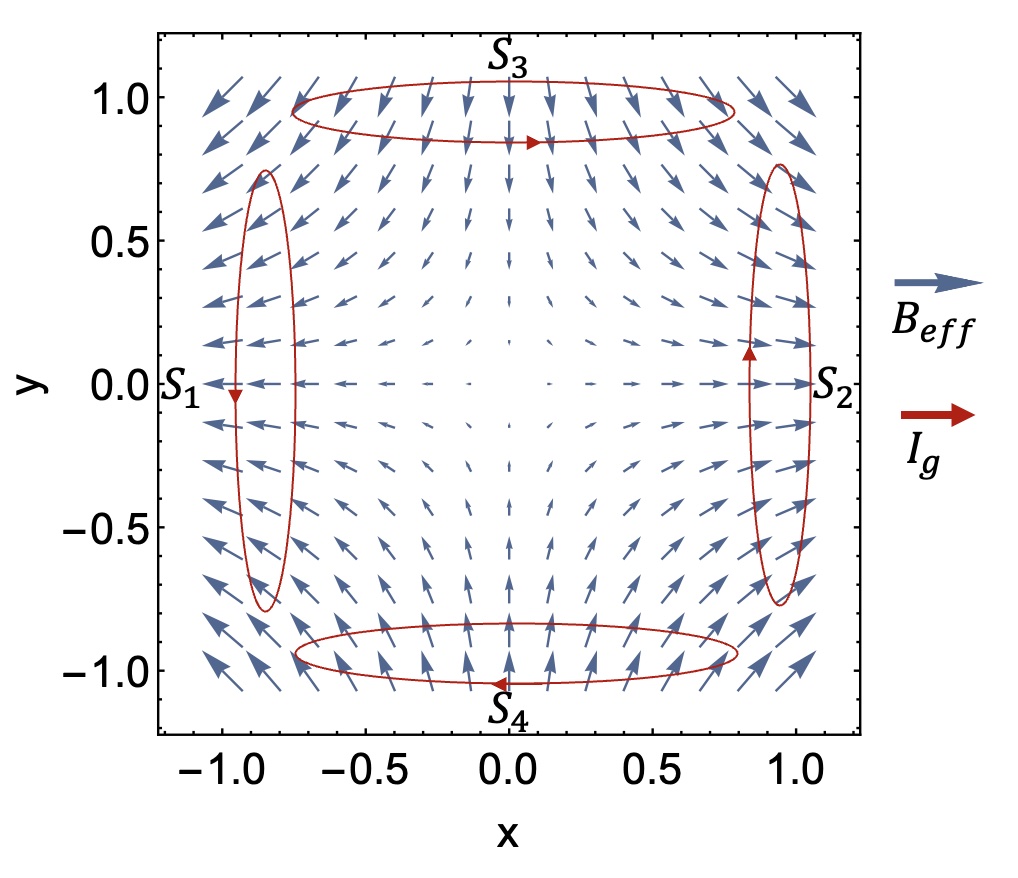}\label{fig.sensor}}
		\subfloat[Gert. effect ($h_\times$).]{\includegraphics[width=0.4\linewidth]{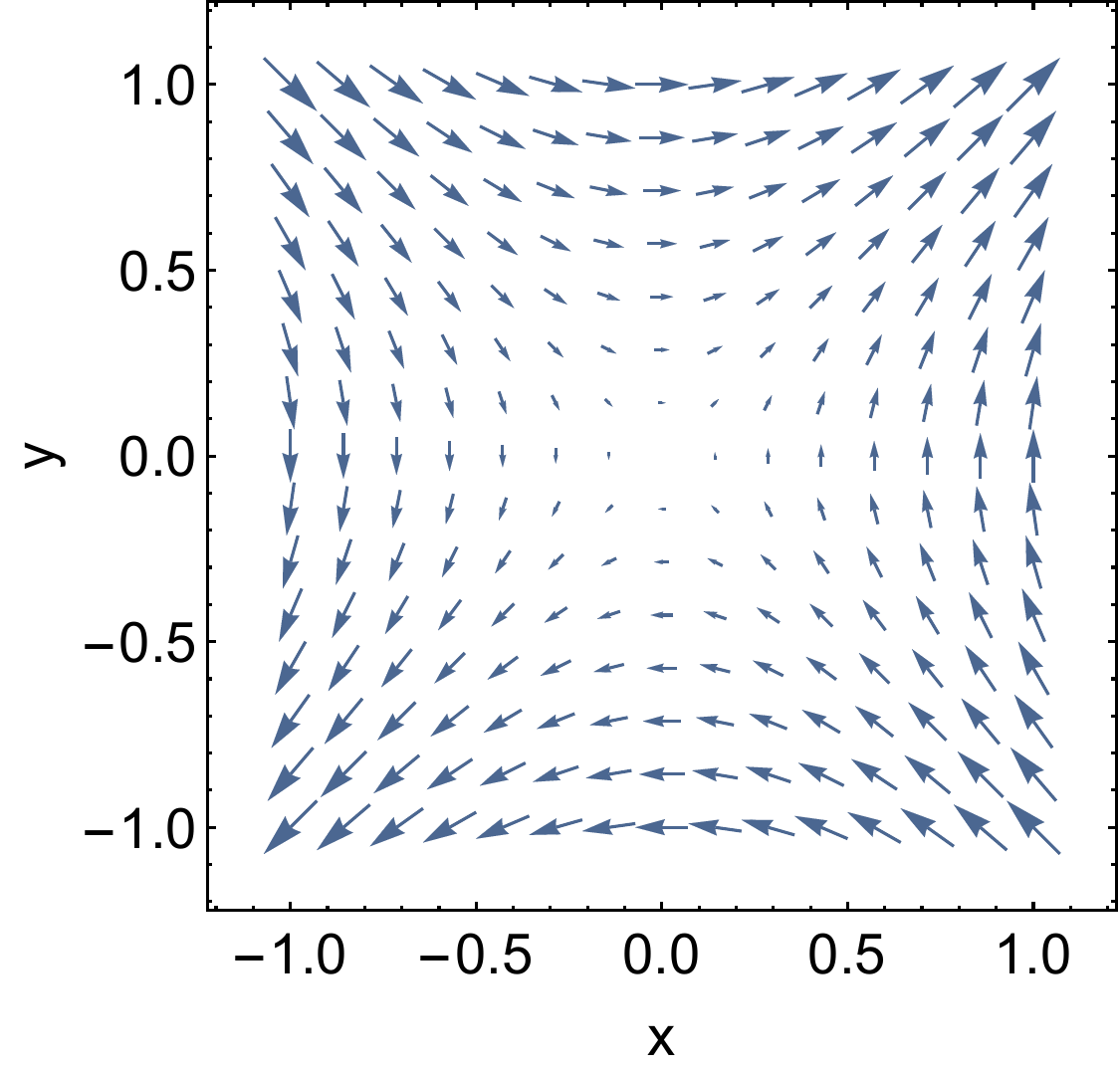}}
	\caption{Spatial distributions for $B_{\rm metric}^i$ (upper panels) and magnetic field caused by Gertsenshtein effect (lower panels) with natural normalization. The length is given in units of $L$.
	In panel (c), we illustrate the GW detection scheme using an array of four sensors $S_1 - S_4$. 
	}
	\label{p112} 
\end{figure}

{\textbf{Detectability.}
\label{sec3}
The effective magnetic fields induced by gravitational waves and experienced by spins—arising from $B_{\rm metric}^i$ and the Gertsenshtein effect—are shown in Figure~\ref{p112}.

Since the dominant Gertsenshtein effect scales with the strength of the background magnetic field, we focus on detecting gravitational waves at high frequencies, above the kilohertz range. Nuclear spins offer a significant advantage over electron spins due to their much longer coherence times. Moreover, under the same magnetic field, nuclear spins can generate a stronger signal from the metric-induced effect. For these reasons, we choose nuclear spins as the sensing medium and target the frequency range from kilohertz to sub-gigahertz. 
Representative experiments that employ nuclear spin-based detection include the CASPEr experiment~\cite{budker2014proposal} that uses $^{129}$Xe for dark matter search and the ARIADNE experiment~\cite{arvanitaki2014resonantly,gkika2024optimization} that uses $^{3}\rm He$ to search for new forces beyond the standard model. In this work, we choose $^{129}\mathrm{Xe}$ as a representative sensing nucleus and adopt parameters similar to those used in the CASPEr-wind Phase-II experiment~\cite{garcon2017cosmic,jackson2020overview}.

The atoms are initially polarized with polarization $p=0.5$ along the z-axis via optical pumping~\cite{garcon2017cosmic} at a spin density $n_s\approx 10^{22} ~{\rm cm}^{-3}$, providing a magnetization $M_z$. We sum up all the GW-induced effects as an effective field $B_{\text{eff}}$. A time-varying gravitational wave with amplitude $B_{\text{eff}}$ and frequency $\omega$, traveling along the {\it z}-axis, drives the spins and induces a resonant response, leading to the development of a transverse magnetization $M_{x,y}$. An example distribution is shown in Fig.~\ref{p112}c, where detectors are located in circled areas. The magnitude of $M_{x,y}$ is
$$
M_{x,y}(t) \approx \frac{1}{2} n_s p \mu_N \gamma_N B_{\mathrm{eff}} T_2\left(e^{-t / T_1}-e^{-t / T_2}\right) \cos (\omega t)
$$
where $\mu_N$ represents the nuclear magnetic moment, the transverse magnetization $M_x(t)$ builds up nearly linearly over time until it reaches the characteristic timescale of coherence time $T_2$, 
and then decays at the longer longitudinal relaxation time $T_1$. 
The magnetization can be detected using a pickup loop oriented perpendicular to the initial spin polarization, which maximizes sensitivity. The transverse magnetization $M_{x, y}$ acts as an effective signal amplifier, quantified by the ratio $\eta_{\mathrm{NMR}}=\mu_0 M_{x, y} / B_{\text {eff }}$. The nuclear spin coherence time $T_2$ is limited by magnetic field inhomogeneity (typically $\sim 0.1 \mathrm{ppm}$ in well-optimized systems). At ultra-low fields (e.g., a few $\mu \mathrm{T}$ ), $T_2$ can extend to $\sim 1000 \mathrm{~s}$, while $\eta_{\mathrm{NMR}}$ reaches amplification factors as high as $10^6$~\cite{jackson2020overview}.

Detection is typically performed in gradiometer mode, which uses two adjacent sensors to suppress common-mode magnetic noise while maintaining high sensitivity to magnetic field gradients.\,\cite{gkika2024optimization,walter2025search}. This approach is particularly well-suited for gravitational wave detection, as gravitational waves inherently produce gradient-like field patterns. 

As illustrated in Fig.\,\ref{fig.sensor}, we consider an experimental setup consisting of four sensors (in this case, pickup coils labeled $S_1$ to $S_4$), placed in the sensing region. $S_1$ and $S_2$ are oriented to detect signals along the {+\it x} and {-\it x} directions, respectively, while $S_3$ and $S_4$ are aligned to measure along the +{\it y} and -{\it y} directions. The GW signal will manifest as a magnetic field $B_g$ and exhibit a quadrupole pattern. The joint detectors at different locations within the magnetic field will exhibit a distinct pattern and can be identified with different gravitational wave modes. 
In contrast, the common-mode magnetic noise $B_{CN}$ in the detection area is usually uniform and can be subtracted. At frequencies below 1 MHz, SQUIDs perform well, while at higher frequencies above 1 MHz, conventional LC circuits can be employed~\cite{bloch2023scalar}. A typical white-noise level of a commercial SQUID is $\delta \Phi_{\mathrm{sq}} \approx 1 \mu \Phi_0 / \sqrt{\mathrm{Hz}}$, where $\Phi_0$ is the quantum of magnetic flux. The discussion of the detection mode and the comparison with a normal NMR mode can be found in Appendix C.
{\textbf{Projected sensitivity.}} We choose the baseline distance $d_{base}=2L$ between $S_1$ and $S_2$ as 20 cm (same for $S_3$ and $S_4$), which can be easily applied in a conventional magnetically shielded chamber, either using $\mu$-metal shielding or superconducting shielding. 
For GWs from 1\,kHz to 0.4\, GHz, the GW wavelength is larger than $d_{base}$, and we can assume the GW is coherent in the detection area, and assume the coherence time of the GW is longer than the $T_2$ of nuclear spins. 
Bosonic fields centered with a Kerr Black Hole can emit monochromatic GWs with a coherence time longer than $T_2$ as Eq. (\ref{srtime}) explains.
The upper limit is constrained by the maximum dc magnetic field around 30\,T.

The total effect of GWs on a spin can be equivalent to an effective magnetic field, which a magnetometer can detect. 
As discussed above, the  effective magnetic field
\begin{equation}
\begin{aligned}
 B^i_{eff}&\sim  B^i_{\rm metric}+B^i_{\rm Gertsenshtein}\\
 &\sim h \left(\left(\frac{\omega^2 L}{2\sqrt{2}\gamma}\right)^2+\left(\frac{L\omega B_0}{\sqrt{2}}\right)^2\right)^{1/2}.
\end{aligned}
\end{equation}
Notice that these two effects have a phase difference of $\pi/2$. Thus, the equivalent sensitivity to GWs  can be estimated by 
\begin{equation}
\begin{aligned}
 \sqrt{S_h}\sim \left(\left(\frac{\omega^2 L}{2\sqrt{2}\gamma}\right)^2+\left(\frac{L\omega B_0}{\sqrt{2}}\right)^2\right)^{-1/2} \sqrt{S_B}
\end{aligned}
\end{equation}
To estimate the GW sensitivity, we take $\sqrt{S_B}$ from Ref.~\cite{jackson2020overview}, where it also gave a limit on the $g_{aNN}$ coupling. The corresponding axion induced pseudo-magnetic field is converted by $B_a[T]\simeq 10^{-7}\times \frac{g_{aNN}[\text{GeV}^{-1}]}{g_N}$, where $g_N$ is Lande factor for $^{129}$Xe. The corresponding power spectrum for the magnetic field is estimated by $\sqrt{S_B}\sim B_a\sqrt{\tau_a}$, where $\tau_a\sim 10^6/f$ is the axion coherence time.
The equivalent strain of GWs can be detected and is shown in Fig. \ref{fig:p1}. We can see that spin-based systems, even without special detector designs, can achieve competitive sensitivity at higher frequencies.

%

\begin{figure}[t]
	\includegraphics[width=0.95\linewidth]{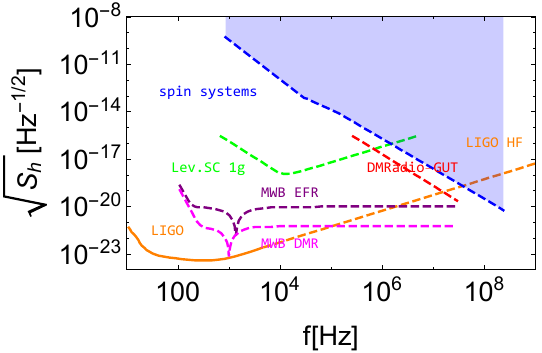}
\caption{Projected GW sensitivity for spin systems (blue). 
For comparison, other proposed sensitivity curves from superconducting levitated detectors~\cite{Carney:2024zzk}, DM-Radio~\cite{DMRadio:2022jfv}, magnetized Weber bar~\cite{Domcke:2024mfu}, and extrapolated LIGO sensitivity to higher frequencies~\cite{Aggarwal:2025noe} are shown in green, red, and orange colors, respectively. }
	\label{fig:p1}
\end{figure}

{\textbf{Conclusion.}}\label{sec4}
In this work, we have revisited the interaction between spins and GWs. By identifying the effect of GWs on spins as an effective magnetic field, we propose using an NMR-based magnetometer to detect GWs.
The effective magnetic field $B^i_{\rm metric}$ that comes from the GW-spin coupling, and the magnetic field induced by the Gertsenshtein effect are studied. We show that the  Gertsenshtein effect dominates, but the $B^i_{\rm metric}$ is non-negligible. 
Using the magnetic sensitivity from Ref.~\cite{jackson2020overview}, the corresponding sensitivity to gravitational waves can reach $\sqrt{S_h}\sim 10^{-20}\text{Hz}^{-1/2}$ at 100 MHz. 
The sensitivity $h_c \propto \frac{\lambda}{L}$ is enhanced for a larger spin-detector separation $L$ under the magnetic field, while $L$ is still below the GW wavelength $\lambda$. For the longer GW wavelengths,
small-sized detectors will have a limited sensitivity, but may benefit from joining an extensive network like GravNet~\cite{afach2021search}. 
In addition to GW detection, we note that magnetometers can also detect the frame-dragging effects due to the Earth's rotation ~\cite{Everitt:2011hp}. We emphasize that magnetometers, especially the levitated ferromagnetic particle, can be used to detect the interaction of gravity and spins, which differs from traditional GW detection based on the inherent length change caused by GWs\cite{fadeev2021gravity,ji2025levitated}.

{\bf Acknowledgments.}
We want to thank Dmitry Budker for the helpful discussions. The authors acknowledge support from the National Natural Science Foundation of China (Grant No. 12447105).

\bibliography{sample}
\newpage
\onecolumngrid

\section{Appendix}

\subsection{Derivation of Gravitation Wave}

We focus on the effect of GWs on spins. The Hamiltonian of Dirac fermions can be obtained from the Dirac equation.  Using the Foldy-Wouthuysen transformation as in Ref.~\cite{Asprea:2020fuy}, the non-relativistic Hamiltonian with spin is: 
\begin{equation}\label{Hspin}
	\begin{aligned}
		H_s=\frac{1}{4}\epsilon^{ijk}h_{0j,i}\Sigma_k+\frac{\epsilon^{ijk}}{4m}\left(h_{00,i}\hat{p}_j-h_{jl,i}\hat{p}^l\right)\Sigma_k\,.
	\end{aligned}	
\end{equation}
The second term describes a gravitational spin-orbit coupling, which leads to the gravitational spin Hall effect, as discussed in Ref. ~\cite{Wang:2023bmd}. In the non-relativistic scenario, the second term is of order $p/m$ and can thus be neglected. In the transverse-traceless (TT) gauge, $h_{0i}=0$; thus, GWs do not affect the spins. However, we emphasize that the laboratory's preferred frame is the Fermi normal frame, which is the local inertial coordinates along a geodesic. In Fermi normal coordinates, the first term can lead to spin precession. There is recent discussion about the validity of using a single laboratory frame across the apparatus~\cite{Bringmann:2023gba}, suggesting that a modification factor may be required depending on the frame's rigidity.

The first term in Eq.(\ref{Hspin}) can be obtained from a gravito-electromagnetic analogy. Neglecting the $\mathcal{O}(c^{-4})$ terms in metric, the spacetime metric has the form, 
\begin{equation}
	\begin{aligned}
 d s^{2}=-\left(1+2\Phi\right)d t^{2}-4(A\cdot d x)d t+\left(1-2\Phi\right)\delta_{i j}d x^{i}d x^{j},
\end{aligned}
\end{equation}
where we have rewrote $\Phi=-h_{00}/2, A_i=-h_{0i}/2$. For a classical particle, the Lagrangian is $L=-m\frac{ds}{dt}$, thus
\begin{equation}
	\begin{aligned}
 L=-m\sqrt{1-v^2}-\frac{m}{\sqrt{1-v^2}}\left(1+v^2\right)\Phi-\frac{2m}{\sqrt{1-v^2}}v\cdot A
\end{aligned}
\end{equation}
Compared to the Lagrangian of a charged particle in the electromagnetic field, the last term will introduce an effective magnetic field ${\rm B}_{\rm metric}=\nabla\times {\rm A}$ with charge $q=2m$.  Analogy to charged particles in a magnetic field,
The Hamiltonian for a particle with angular momentum $\rm \Sigma/2$ is 
\begin{equation}
	\begin{aligned}
 H_s&=\frac{q}{2m}(\nabla\times {\rm A})\cdot {\rm \frac{\rm\Sigma}{2}}\\
 &=\frac{1}{4}\epsilon^{ijk}h_{0j,i}\Sigma_k\,.
\end{aligned}
\end{equation}

To demonstrate gauge invariance, we will compare the expectation value of the spin operator in two frames and derive the Hamiltonian in Fermi normal coordinates.
In Fermi normal coordinates, the expectation value of the spin operator is
\begin{equation}
	\begin{aligned}	\langle\bar{\psi}\left|\Sigma^i\right|\psi\rangle_{FN}&=\frac{i \epsilon^{ijk}}{2}\langle\bar{\psi}\left|\gamma_j\gamma_k\right|\psi\rangle_{FN}\\
		&=\frac{i\epsilon^{ijk}}{2}\langle\bar{\psi}\left|\gamma_\mu\gamma_\nu\right|\psi\rangle_{TT}\frac{\partial x^\mu_{TT}}{\partial x_j}\frac{\partial x^\nu_{TT}}{\partial x_k},\\
	\end{aligned}
\end{equation}
where the coordinate transformations are
\begin{equation}
	\begin{aligned}
		t_{TT}&\simeq t+\frac{\omega}{4\sqrt{2}}\left(x^{2}-y^{2}\right)h_{+}\sin\theta_++\frac{\omega xy}{2\sqrt{2}} h_{\times}\sin\theta_\times,\\
		x_{{TT}}&\simeq x-\frac{h_{+}x}{2\sqrt{2}}(\cos\theta_++\omega z\sin\theta_+)-\frac{h_{\times}y}{2\sqrt{2}}(\cos\theta_\times+\omega z\sin\theta_\times),\\
		y_{TT}&\simeq y+\frac{h_{+}y}{2\sqrt{2}}(\cos\theta_++\omega z\sin\theta_+)-\frac{h_{\times}x}{2\sqrt{2}}(\cos\theta_\times+\omega z\sin\theta_\times),\\
		z_{\mathrm{TT}}&\simeq z+\frac{\omega}{4\sqrt{2}}\left(x^{2}-y^{2}\right)h_{+}\sin\theta_++\frac{\omega xy}{2\sqrt{2}} h_{\times}\sin\theta_\times.
	\end{aligned}
\end{equation}
 We have assumed the GWs are along the z direction, and $\theta_+=\omega t\,, \theta_\times=\omega t+\alpha$ are the phases for the two modes of GWs. Supposing $\langle\bar{\psi}\left|\Sigma^1\right|\psi\rangle_{TT}=\langle\bar{\psi}\left|\Sigma^2\right|\psi\rangle_{TT}=0,\langle\bar{\psi}\left|\Sigma^3\right|\psi\rangle_{TT}=1$ for simplicity, after some calculations, we obtain
\begin{align}\label{spin11}
 \langle\bar{\psi}\left|\vec{\Sigma}\right|\psi\rangle_{FN}=\hat{e}_z+\frac{\omega h_+}{2\sqrt{2}}\sin\theta_+(x\hat{e}_x-y\hat{e}_y)+\frac{\omega h_\times}{2\sqrt{2}}\sin\theta_\times(y\hat{e}_x+x\hat{e}_y). 
 \end{align}
Thus, 
\begin{equation}
	\begin{aligned}
\frac{d}{dt}\langle\bar{\psi}\left|\vec{\Sigma}\right|\psi\rangle_{FN}&=\frac{\omega^2 h_+}{2\sqrt{2}}\cos\theta_+(x\hat{e}_x-y\hat{e}_y)+\frac{\omega^2 h_\times}{2\sqrt{2}}\cos\theta_\times(y\hat{e}_x+x\hat{e}_y)\\
&=i\langle\bar{\psi}\left|\left[\hat{H}_{FN},\vec{\Sigma}\right]\right|\psi\rangle_{FN}\\
\text{with}\quad \hat{H}_{FN}&=\frac{\omega^2 h_+}{4\sqrt{2}}\cos\theta_+(y\Sigma^1+x\Sigma^2)+\frac{\omega^2 h_\times}{4\sqrt{2}}\cos\theta_\times(-x\Sigma^1+y\Sigma^2).
\end{aligned}
\end{equation}
We have shown that $\hat{H}_{FN}$ fulfills the Heisenberg equation. The Hamiltonian $\hat{H}_{FN}$ can also derive from Eq.(\ref{fnhij}) and (\ref{Hspin}). 

\subsection{Expression for the induced electromagnetic fields}

{\color{black}Considering an static magnetic field ${\bf B}_0=B_0\hat{z}$, we have tensor $h$ as in eq.(\ref{eqGWS}) 
with $\theta=0$}, then the effective magnetic fields are
\begin{equation}
\begin{aligned}
B_{\rm grav.mag.}^1&=\frac{B_0}{12\sqrt{2}} (h_\times y \omega ^2 z \cos (\alpha +t \omega )\\
&+h_+ x \omega ^2 z \cos (t \omega ))\\
 B_{\rm grav.mag.}^2&=\frac{B_0}{12\sqrt{2}} (h_\times x \omega ^2 z \cos (\alpha +t \omega )\\
 &-h_+ y \omega ^2 z \cos (t \omega ))\\
 B_{\rm grav.mag.}^3&=\frac{B_0}{4\sqrt{2}} (2 h_\times \omega ^2 xy \cos (\alpha +t \omega )\\
 &+h_+ \omega ^2 (x^2-y^2) \cos (t \omega ))
\end{aligned}
\end{equation}
and
\begin{equation}
\begin{aligned}
 B^1_{\rm metric}&=\frac{m\omega^2}{2\sqrt{2}e}(h_\times x \cos (\alpha +t \omega )\\
 &-h_+ y \cos (t \omega ))\\
 B^2_{\rm metric}&=-\frac{m\omega^2}{2\sqrt{2}e}(h_\times y \cos (\alpha +t \omega )\\
 &+h_+ x \cos (t \omega ))\\
 B^3_{\rm metric}&=0.
\end{aligned}
\end{equation}

In addition to the external magnetic field applied to a spin, the induced magnetic field due to the Gertsenshtein effect should not be neglected. As we show next, the Gertsenshtein effect can be significant. It derives from Maxwell's equations $\nabla_{\nu}F^{\mu\nu}=0$. To first order in $h_{\mu\nu}$, the Maxwell equations are
 \begin{align}
	\partial_{\nu}F^{\mu\nu}+\partial_{\nu}\left(\frac{1}{2}h F^{\mu\nu}+h_{~\alpha}^{\nu}F^{\alpha\mu}-h_{~\alpha}^{\mu}F^{\alpha\nu}\right)=0.
\end{align}
Writing $E^i=F^{0i},F^{ij}=\epsilon^{ijk}B^k$, we have
\begin{equation}
	\partial_{i}\left(E^{i}-h^{0}{}_{j}\varepsilon^{j i k}B^{k}\right)=0.
\end{equation}
{\color{black}Then the electric field induced by the magnetic field is}
\begin{equation}
	E^{i}=\varepsilon^{i j k}h_{\;\;j}^{0}B^{k}.
\end{equation}
Assuming the background
magnetic field ${\bf B_0}$ is in the $\hat{z}$ direction and GWs take the form in Eq.~\ref{eqGWS} with $\theta=0$, 
then
\begin{equation}
	\begin{aligned}
		E^1&=-\omega ^2 z (h_+ y \cos (\omega t)+h_{\times} x \cos (\alpha +\omega t))\frac{B_0}{\sqrt{2}},\\
		E^2&=-\omega ^2 z (h_\times y \cos (\alpha +\omega t)+h_+ x \cos (\omega t))\frac{B_0}{\sqrt{2}},\\
		E^3&=0;\\
		B^1&=\left(x\omega h_+\sin(\omega t)+y\omega h_\times\sin(\omega t+\alpha)\right)\frac{B_0}{\sqrt{2}},\\
		B^2&=\left(-y\omega h_+\sin(\omega t)+x\omega h_\times\sin(\omega t+\alpha)\right)\frac{B_0}{\sqrt{2}},\\
		B^3&=B_0.
	\end{aligned}
\end{equation}

\subsection{Comparison with the original CASPEr dark matter detection geometry}

We compare the detection scheme shown in Fig.~2(c) with the conventional NMR-based approach employed in the CASPEr experiment, where the precession of a dipole moment is detected with a pickup coil. In both cases, we consider only the component of the magnetization induced by gravitational waves or dark matter that is transverse to the spin polarization direction. For sensor $S_2$, we approximate the induced magnetization as that of a plate with diameter $L$ and thickness $L/2$, where the magnetization increases gradually from zero at $x=0$ to $M$ at $x=L/2$. The resulting flux is about one-half of that produced by a uniformly magnetized plate of the same geometry with magnetization $M$. Other sensors can be similarly modeled as plates with gradually varying magnetization, differing only in the orientation of their magnetization vectors. Considering sensors $S_1$ and $S_2$ together, the effective flux is approximately equivalent to that of a single plate of thickness $L/2$ and diameter $L$. Including $S_3$ and $S_4$ would provide additional signal, but we do not consider them here because their magnetized regions would overlap with those of $S_1$ and $S_2$, leading to double counting. 

For comparison, a conventional pickup coil in CASPEr is effectively coupled to a plate of length $L$ and diameter $L$, yielding a larger flux due to the reduced demagnetizing field. Using Aharoni’s method~\cite{aharoni1998demagnetizing}, this flux is estimated to be about 1.39 times that of the thinner uniformly magnetized plate. Additional corrections, such as those arising from the non-uniform magnetization profile, may further modify the result, but we estimate the total discrepancy to be within a factor of two. Therefore, we directly adopt the sensitivity estimates from the CASPEr experiment for our comparison.

\end{document}